\documentclass[prl,twocolumn,floatfix,superscriptaddress,notitlepage]{revtex4-1}

\usepackage{amssymb,amsmath,amstext}                
\usepackage{graphicx}                                               
\usepackage{epstopdf}                                               
\usepackage{color}       
\usepackage{bm}
\usepackage{appendix}                                              
\usepackage[utf8]{inputenc}
\usepackage{bbold}
\usepackage{bbm}
\usepackage{ulem}
\usepackage{latexsym}
\usepackage[colorlinks=true,citecolor=blue,linkcolor=magenta]{hyperref}

\begin{document}
\title{Symmetry breaking bias and the dynamics of a quantum phase transition}

\author{Marek M. Rams}
\affiliation{Jagiellonian University, Marian Smoluchowski Institute of Physics,  \L{}ojasiewicza 11, 30-348 Krak\'ow, Poland}

\author{Jacek Dziarmaga}
\affiliation{Jagiellonian University, Marian Smoluchowski Institute of Physics, \L{}ojasiewicza 11, 30-348 Krak\'ow, Poland}

\author{Wojciech H. Zurek}
\affiliation{Theory Division, Los Alamos National Laboratory, Los Alamos, New Mexico 87545, USA}

\begin{abstract}
The Kibble-Zurek mechanism predicts the formation of topological defects and other excitations that quantify how much a quantum system driven across a quantum critical point fails to be adiabatic. We point out that, thanks to the divergent linear susceptibility at the critical point, even a tiny symmetry breaking bias can restore the adiabaticity. The minimal required bias scales like $\tau_Q^{-\beta\delta/(1+z\nu)}$, where $\beta,\delta,z,\nu$ are the critical exponents and $\tau_Q$ is a quench time. We test this prediction by DMRG simulations of the quantum Ising chain. It is directly applicable to the recent emulation of quantum phase transition dynamics in the Ising chain with ultracold Rydberg atoms.
\end{abstract}

\maketitle


In this paper we investigate the interplay of the Kibble-Zurek (KZ) theory of the dynamics of symmetry-breaking quantum phase transitions \cite{K,Z,CZ} with the extreme sensitivity of a quantum critical system to a perturbation by a symmetry breaking bias field. 

There are two complementary motivations of our study. The first one is more fundamental (but more distant). It stems from the discussion textbooks usually offer to justify symmetry breaking phase transition -- see e.g.,~Ref.~\onlinecite{Goldenfeld}. As is noted there, when the relevant thermodynamic potential is symmetric, there is no reason for the system to settle in a particular broken symmetry post-transition state. This leads to a conceptual difficulty -- why does the symmetry break? In textbooks it is usually addressed by introducing an external bias that favors a particular post-transition state (e.g., a magnetic field that biases spins during a ferromagnetic phase transition). That field is eventually allowed to vanish, and if the thermodynamic limit is attained before the bias disappears, one ends up with a broken symmetry phase – same symmetry broken in the same way throughout the whole infinite volume. It is however important to attain the thermodynamic limit (infinite system size) prior to turning off the bias. Similar considerations apply to quantum phase transitions. For instance, in the quantum Ising model in a transverse field the ground state is degenerate, spanned by the symmetric superposition of the two obvious broken symmetry ferromagnetic states. 

\begin{figure} [t]
\begin{center}
 \includegraphics[width=\columnwidth]{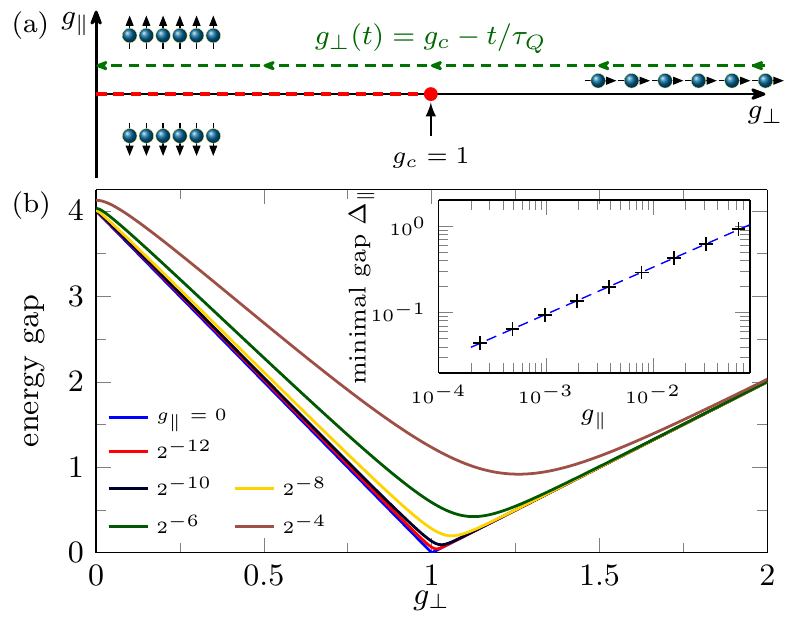}
\end{center}
  \caption{In (a), phase diagram for the ferromagnetic Ising model in Eq.~\eqref{hamiltonian}. Transverse field $g_\perp$ is slowly driven from a paramagnetic to a ferromagnetic phase (dashed green line) close to the continuous Ising critical point (red dot). Red dashed line indicates first order transition between differently oriented ferromagnetic states. In (b), bias $g_{\|}$ opens up a gap near the critical point. Excitation energy at zero momentum is calculated using uniform matrix product states ansatz directly in the thermodynamic limit~\cite{utdvp_review, MPSexcitations}. In the inset we show the scaling of minimal gap as a function of the bias. 
  For the presented range we fit $\Delta_{\|} \propto g_{\|}^{0.545}$, where the exponent is close to the expected $z \nu_{\|} = 8/15 \approx 0.5333$ and leaning towards it for smaller $g_{\|}$.} 
   \label{fig:KZ}
\end{figure}
In contrast to that ``textbook’’ motivation (which deals with a sequence of equilibria) one is often (e.g., in the quantum information processing context) interested in reaching states that are similarly ordered, but attaining it reasonably quickly. Thus, the timescale is of the essence. One can expect that rapid transitions will introduce disorder – excitations and defects. This is indeed the conclusion of a KZ mechanism (KZM) for quantum phase transitions. A second motivation for our study is therefore to consider whether an external bias – longitudinal field in the transverse field quantum Ising model – may effectively suppress the formation of such defects, and hence, be used to evade spurious excitations. This is very much in the spirit of the ``shortcuts to adiabaticity’’ \cite{shortcuts}. It is also of interest because – in the experiments that test the dynamics of quantum phase transitions (see e.g., Ref.~\onlinecite{Lukin18}) -- one may be concerned whether the symmetry breaking was truly spontaneous or whether it was in fact ``suggested’’ by external perturbations. 

The classical KZM was recognized as a general theory of the dynamics of phase transitions and verified by numerical simulations \cite{KZnum} and laboratory experiments in various condensed matter systems \cite{KZexp}. More recently, it was generalized to quantum phase transitions \cite{Polkovnikov2005,QKZ1,QKZ2,d2005,d2010,p2011}. Theoretical advances \cite{QKZteor} and experimental tests \cite{QKZexp,deMarco2,Lukin18} followed. The recent experiment, with Rydberg atoms \cite{Lukin18}, is an accurate quantum simulation of the exact solution in the quantum Ising chain \cite{d2005}.

In quantum KZM a system initially prepared in its ground state is smoothly ramped across a critical point to the other side of the phase transition. A distance from the critical point -- measured by a dimensionless parameter $\epsilon$ -- can be linearized close to the criticality as
\begin{equation}
\epsilon(t)=\frac{t}{\tau_Q}.  
\label{epsilont}
\end{equation}
Here $\tau_Q$ is the quench time. Initially, far from the transition, the evolution is adiabatic and the system follows its adiabatic ground state. The adiabaticity fails at $-\hat t$ when the rate of the transition, $\dot\epsilon/\epsilon=1/|t|$, becomes faster than the gap between the ground state and the first relevant excited state. The gap closes like $\Delta\propto |\epsilon|^{z\nu}$, where $z$ and $\nu$ are the dynamical and correlation length exponents, respectively. From the equation $1/|t| \propto |t/\tau_Q|^{z\nu}$ we obtain $\hat t \propto \tau_Q^{z\nu/(1+z\nu)}$ and a corresponding $\hat\epsilon=\hat t/\tau_Q \propto \tau_Q^{1/(1+z\nu)}$. In naive, but surprisingly successful adiabatic-impulse approximation, the ground state at $-\hat\epsilon$, with a correlation length
\begin{equation}
\hat\xi \propto \tau_Q^{\nu/(1+z\nu)},   
\end{equation}
remains the state of the system until $+\hat t$, when the evolution becomes adiabatic again. More accurate ``sonic horizon'' view of this process yields the same scalings. In this way $\hat\xi$ becomes imprinted on the initial state for the final adiabatic stage of the evolution. KZM can be characterized by a length and timescale, $\hat\xi$ and $\hat t\propto \hat\xi^z$, both diverging when $\tau_Q\to\infty$.

Due to the diverging linear susceptibility at the critical point, even a tiny symmetry breaking bias $g_{\|}$ can derail the KZM. It induces a nonzero magnetization 
$M_{\|}\equiv \langle \sigma^z\rangle \propto g_{\|}^{1/\delta}$ that in turn implies a nonzero gap 
$\Delta_{\|} \propto g_{\|}^{z\nu/\beta\delta}$
and a finite correlation length 
$\xi_{\|}\propto g_{\|}^{-\nu/\beta\delta}$. Here $\delta$ and $\beta$ are critical exponents, where we follow the standard convention, see e.g., Ref.~\onlinecite{ContinentinoBook}. 
In the Ising universality class $\nu/\beta\delta= \nu_\| = 8/15$ is the famous Zamolodchikov's correlation length exponent \cite{Zamolodchikov}.
In presence of longitudinal field, and at odds with the KZM assumptions, neither the gap closes nor the correlation length diverges at the critical point. Therefore, we expect that when the bias is strong enough, i.e., for $\xi_{\|}\ll\hat\xi$ or, equivalently,
\begin{equation}
\tau_Q\gg g_{\|}^{-(1+z\nu)/\beta\delta},
\label{adiabcond}
\end{equation}
then the ramp (\ref{epsilont}) becomes adiabatic. In the opposite regime, for small $\tau_Q$, KZM should work as usual because $-\hat\epsilon$, when the state freezes out, is too large for the tiny bias to have anything more than a perturbative effect.

In the following we test this prediction with DMRG simulations in the transverse field quantum Ising chain. The chain is not only the standard theoretical workhorse -- exactly solvable without the bias \cite{d2005} -- but also the subject of a recent experiment \cite{Lukin18} testing KZM in a chain of $51$ Rydberg atoms emulating $51$ spins. In this setup the bias is a difference between a contribution from the Rydberg blockade and one due to laser detuning. A fine-tuning is necessary to nullify the net bias. One of the aims of this work is to estimate how accurate the fine-tuning has to be for the bias to become negligible. Already the condition (\ref{adiabcond}) provides a scaling relation between the minimal adiabatic quench time and the bias. With the Ising exponents ($z=\nu=1$, $\beta=1/8$, $\delta=15$) it reads 
$ 
\tau_Q\gg g_{\|}^{-16/15}. 
$ 
Our simulations substantiate the relation with a numerical prefactor that makes it directly applicable to the experimental setup \cite{Lukin18}.

The bias may be an obstacle for testing KZM, but for the adiabatic quantum state preparation it is the KZM itself that is an obstacle preventing adiabatic passage across quantum criticality. The importance of this roadblock was recognized by leading experimental groups, see e.g.,~Ref.~\onlinecite{GreinerHubbard}, and multiple strategies -- under an umbrella name ``shortcut to adiabaticity'' -- were devised in order to bypass it, see a recent review in Ref.~\onlinecite{shortcuts}.
Their design often requires an exact solution of the model that is either not possible, or, if is is possible, the optimal strategy is not easy to implement \cite{StoAPRL}. In contrast, in evolution across a symmetry-breaking phase transition, the symmetry breaking bias is a robust shortcut to adiabaticity.

{\it Excitations.---}
We consider the quantum Ising chain in both transverse and longitudinal magnetic fields: 
\begin{equation} 
\label{hamiltonian}
H = - \sum_{n} \left[\sigma^x_n \sigma^x_{n+1} + g_{\perp} \sigma^z_{n} + g_{\|} \sigma^x_{n} \right],
\end{equation}
It has an isolated critical point at $(g_{\perp} ,g_{\|}) =(1,0)$ between the paramagnetic and ferromagnetic phases, see Fig.~\ref{fig:KZ}(a). We ramp the transverse field as 
\begin{equation}
g_{\perp}(t) \simeq 1-\frac{t}{\tau_Q},
\label{ramp}
\end{equation}
from $+\infty$ to $0$, i.e., from an initial ground state deep in the paramagnetic phase, across the
critical $g_\perp=1$, to the ferromagnetic phase \cite{footnote1}. Here we apply also a small bias: 
\begin{equation}
g_{\|}(t) = \rm{const}.
\end{equation}
The question is how strong $g_{\|}$ has to be in order to bypass the critical point $(g_{\perp},g_{\|}) =(1,0)$ adiabatically.

The case of $g_{\|}=0$ is exactly solvable \cite{d2005}. For slow enough quenches the final density of kinks is
\begin{equation}
n_{ex}(g_\perp=0) = \frac{1}{2\pi\sqrt{2\tau_Q}}\propto \hat\xi^{-1}.
\label{nsqrttauq}
\end{equation}
Moreover,
during the KZ stage of the evolution, 
between $\pm\hat t$,
the density of quasiparticle excitations satisfies a KZ scaling hypothesis \cite{KZscaling,Francuzetal}
\begin{equation}
n_{ex}(t) = \hat\xi^{-1} F_{n_{ex}}\left(t/\hat\xi^z\right),    
\end{equation}
where $F_{n_{ex}}$ is a nonuniversal scaling function. In other words, when a scaled density $\hat\xi n_{ex}(t)$ is plotted as a function of scaled time $s=t/\hat\xi^z$ for different $\tau_Q$, the plots collapse to the common scaling function $F_{n_{ex}}(s)$. 

\begin{figure} [t]
\begin{center}
 \includegraphics[width= \columnwidth]{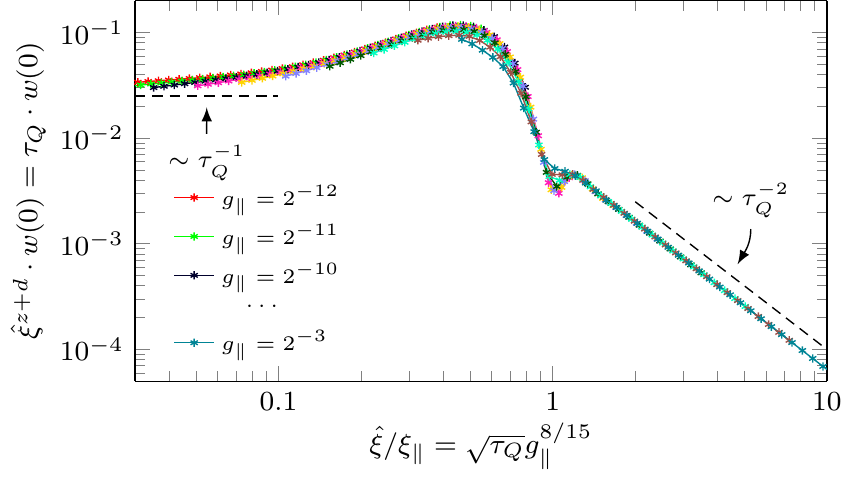}
\end{center}
  \caption{
  Scaled density of excitation energy $\hat\xi^{d+z} w(0)=\tau_Q w(0)$ in the function of $\hat\xi/\xi_{\|}=\sqrt{\tau_Q}g_{\|}^{8/15}$ for different longitudinal biases $g_{\|}$, see Eq.~\eqref{scalingw}. Comparison of the two length scales $\sqrt{\tau_Q}g_{\|}^{8/15}\approx 1$ marks a crossover from the KZ regime with $w(0)\propto \tau_Q^{-1}$
  to an adiabatic regime with $w(0)\propto \tau_Q^{-2}$. Dashed lines provide the guidance for an eye for the expected scaling (slope) on the rescaled plot.
  } 
   \label{fig:w0}
\end{figure}

With the bias the model is not exactly solvable and the quasiparticles become ill-defined. We must instead rely on density of excitation energy $w$ \cite{Polkovnikov2005}. During the KZ stage, we expect it to satisfy a scaling hypothesis \cite{Francuzetal}:
\begin{equation}
w(t) = \hat\xi^{-d-z} F_w\left(t/\hat\xi^z,\hat\xi/\xi_{\|}\right).    
\label{scalingw}
\end{equation}
We test it by numerical simulations with uniform matrix product states employing the time-dependent variational principle \cite{tdvp,utdvp_review}. Its inverse-free formulations in the thermodynamic limit allows us to investigate long evolution times.
Figure \ref{fig:w0} shows a scaled energy density at the critical point, $\hat\xi^{d+z} w(0)=\tau_Q w(0)$,
in function of $\hat\xi/\xi_{\|}=\sqrt{\tau_Q}g_{\|}^{8/15}$. In accordance with the hypothesis (\ref{scalingw}), the plots for different biases collapse to a common scaling function.
Furthermore, near $\sqrt{\tau_Q}g_{\|}^{8/15} \sim 1$ we can clearly see a crossover from the 
KZ regime, where $w\propto \hat\xi^{-d-z}=\tau_Q^{-1}$, to the adiabatic regime where the decay with $\tau_Q$ is faster: $w(0)\propto \tau_Q^{-2}g_{\|}^{-16/15}$.

In order to explain the adiabatic power law, note first that the gap near the critical $g_\perp=1$ scales like $\Delta_{\|}\propto g_{\|}^{z\nu/\beta\delta}$, see Fig. \ref{fig:KZ}(b). This is the ``closest approach'' gap at the center of the Landau-Zener anticrossing. The whole anticrossing takes place between $\pm\epsilon_{\|}$.
Here $\epsilon_{\|}$ is a half-width of the anticrossing where $\epsilon_{\|}^{-\nu}\propto \xi_{\|}$.
Therefore, the whole anticrossing takes time $\tau_Q^{\|}\propto\tau_Q\epsilon_{\|}$ and
the excitation probability at the anti-crossing center is $p\propto \left (\Delta_{\|}\tau_Q^{\|} \right)^{-2}$.
Consequently, the excitation energy is $p\Delta_{\|}\propto\Delta_{\|}^{-1}{\tau_Q^{\|}}^{-2}$.
The energy multiplied by density of available excitations, $\xi_{\|}^{-1}$,
becomes the excitation energy density 
$w(0)\propto \xi_{\|}^{-1}\Delta_{\|}^{-1}{\tau_Q^{\|}}^{-2}=
\tau_Q^{-2}g_{\|}^{(\nu-z\nu-2)/\beta\delta}=\tau_Q^{-2}g_{\|}^{-16/15}$.

\begin{figure} [t]
\begin{center}
  \includegraphics[width=\columnwidth]{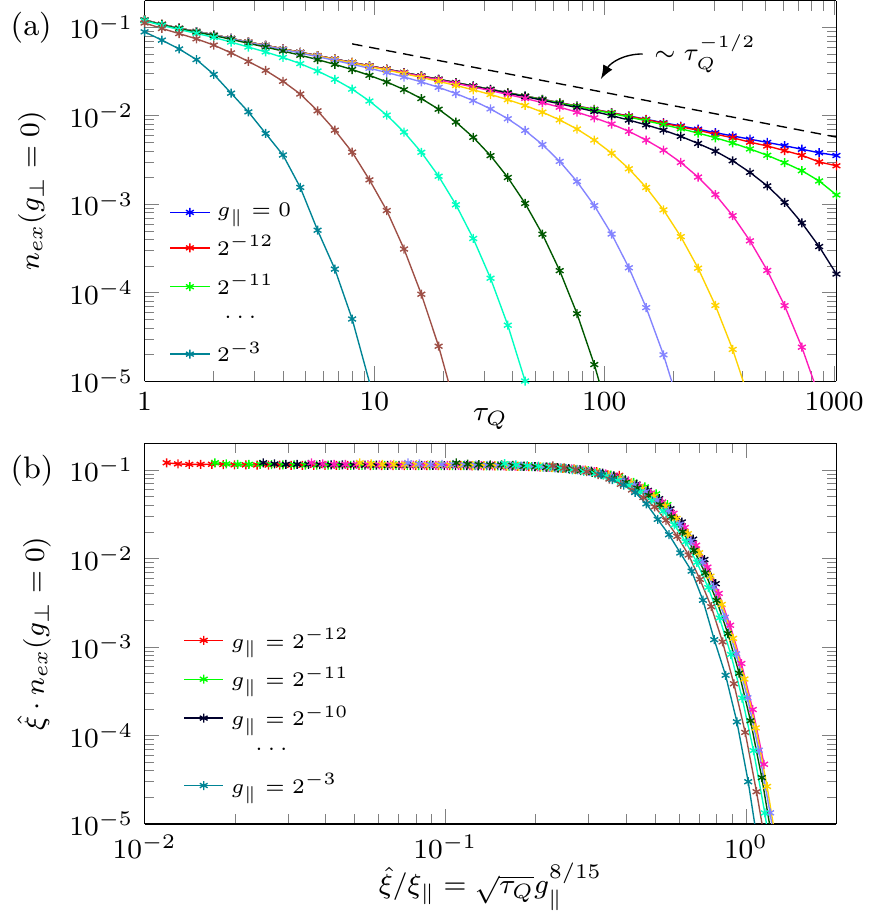}
  \end{center}
  \caption{
  In (a), 
  the final density of kinks at the end of the ramp, $g_\perp = 0$, in function of $\tau_Q$. 
  Different colors correspond to different longitudinal biases $g_{\|}$.
  In (b),
  the scaled final density of kinks $\hat\xi n_{ex}(g_\perp=0)$ in function of the ratio $\hat\xi / \xi_{\|}$.
  There is a clear crossover from the KZ scaling, $n_{ex}(g_\perp=0)\sim\hat\xi^{-1}=\tau_Q^{-1/2}$, to an exponential decay in the adiabatic regime.} 
   \label{fig:g0kinks}
\end{figure}

The energy scaling hypothesis (\ref{scalingw}) demonstrates the crossover from the KZ to the adiabatic regime, but it is not applicable when a constant $g_\perp=0$ is considered instead of a constant 
$t/\hat\xi^z$, see Ref.~\onlinecite{Francuzetal}. At the final $g_\perp=0$, for a weak bias the energy density becomes approximately equal to the density of kinks in the final ferromagnetic state, and in the experiments it is the kinks, rather than the energy, that are directly counted \cite{Lukin18}. Therefore, in Fig. \ref{fig:g0kinks}a we show the final density of kinks as a function of the quench time $\tau_Q$ for different bias fields $g_{\|}$. For any given bias we find that when $\tau_Q$ is small enough, then the kink density scales as $\hat\xi^{-1}\propto\tau_Q^{-1/2}$ like in the unbiased quantum Ising chain (\ref{nsqrttauq}). For longer $\tau_Q$ the dependence crosses over to an exponential decay and the crossover value of $\tau_Q$ decreases with the bias. These observations are corroborated by Fig. \ref{fig:g0kinks}(b) showing the scaled final density of kinks, $\hat\xi n_{ex}(g_\perp=0)$, in function of $\hat\xi/\xi_{\|}$. The plots for different $\tau_Q$ approximately collapse. The collapse improves for weaker biases when the kink density can be identified with the excitation energy more closely. In this limit the exponential decay in the adiabatic regime can be fitted~\cite{footnote1} with
\begin{equation}
    n_{ex}(g_\perp=0)\approx A \tau_Q^{-1/2} e^{-a\tau_Q g_{\|}^{16/15}},
    \label{nexp}
\end{equation}
where $A\approx0.34, a\approx6.89$. The crossover to the adiabatic regime happens for $\tau_Q\approx a^{-1}g_{\|}^{-16/15} \approx 0.145 g_{\|}^{-16/15}$.

\begin{figure} [t]
\begin{center}
  \includegraphics[width=\columnwidth]{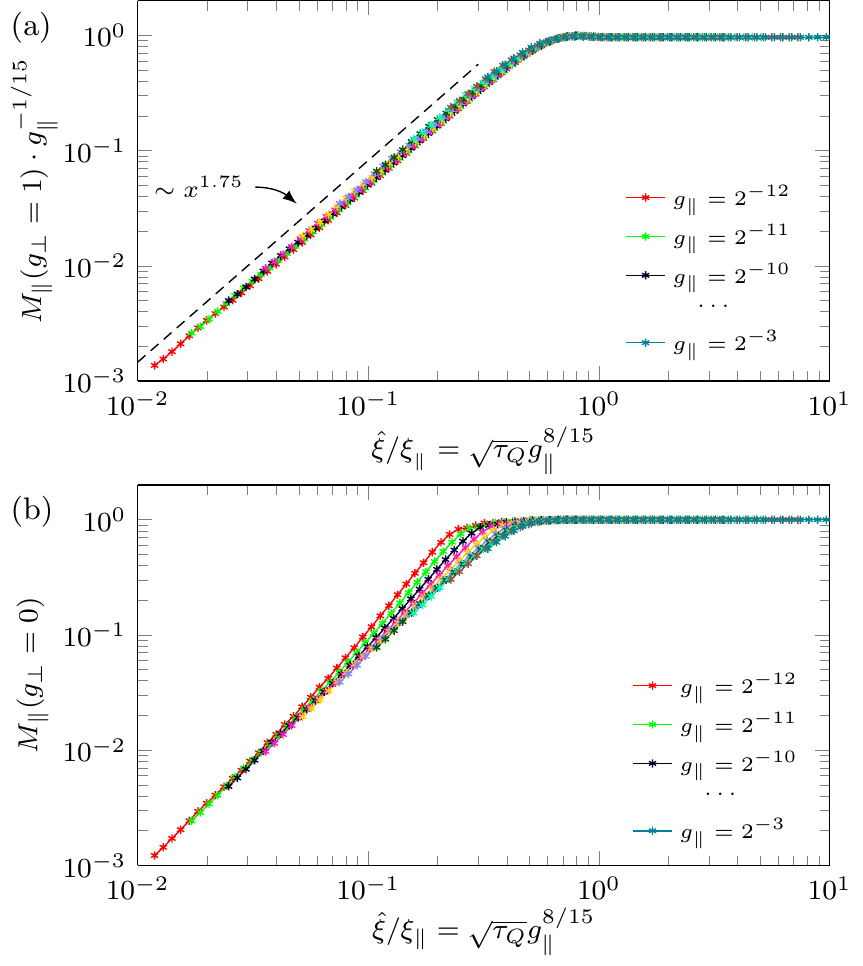}
\end{center}
  \caption{
  In (a), we show
  scaled longitudinal magnetization $g_{\|}^{-1/\delta}M_{\|}(0)$ in function of $\hat\xi/\xi_{\|}$ for different biases, see Eq.~\eqref{Mscaling}. The plots collapse to a common scaling function $F_M\left(0,\hat\xi/\xi_{\|}\right)$,
  in agreement with the dynamical scaling hypothesis.
  In the KZ regime, for relatively slow $\tau_Q$,
  the scaling function is a power law with an exponent close to $1.75$ (dashed line).
  In (b), we show magnetization at the end of the quench
  $M_{\|}(g_\perp = 0)$.
  } 
   \label{fig:ML}
\end{figure}

{\it Longitudinal magnetization.---}
A new feature introduced by the bias is a nonzero longitudinal magnetization. In the ferromagnetic phase, $g_\perp<1$, an infinitesimal bias is enough to induce a spontaneous magnetization removing degeneracy between two ferromagnetic ground states, but the magnetization is strong already at the critical $g_{\perp}=1$, where $M_{\|}\propto g_{\|}^{1/\delta}$. This power law motivates a dynamical scaling hypothesis,
\begin{equation}
    M_{\|}(t) = g_{\|}^{1/\delta} F_M\left( t/\hat\xi^z , \hat\xi/\xi_{\|} \right),
    \label{Mscaling}
\end{equation}
during the diabatic stage between $\pm\hat t$. In Fig. \ref{fig:ML}(a) we show scaled $g_{\|}^{-1/\delta} M_{\|}(0)$ in function of $\hat\xi/\xi_{\|}$ for different biases and find that the plots indeed collapse to a common scaling function $F_M\left( 0 , \hat\xi/\xi_{\|} \right)$.
As for the excitation energy, we can clearly distinguish two regimes. 
In the adiabatic one for relatively slow transitions,
the magnetization follows its static value $\propto g_{\|}^{1/\delta}$ independent of $\tau_Q$.
In the KZ regime for relatively fast transitions,
the collapsed scaling function is a power law that follows from KZM.

In KZM the state freezes out at $-\hat\epsilon\propto -\tau_Q^{-1/(1+z\nu)}$, where the linear susceptibility is $\chi\propto\hat\epsilon^{-\gamma}$ with $\gamma$ being the susceptibility exponent. Its magnetization freezes as
\begin{equation}
  M_{\|}\propto 
  g_{\|}\chi \propto 
  g_{\|} \tau_Q^{\gamma/(1+z\nu)} = 
  g_{\|} \tau_Q^{7/8}. 
  \label{Mpert}
\end{equation}
Here we used $\gamma=7/4$ for the Ising model. Therefore, given the scaling relation $\gamma=\beta(\delta-1)$, in the KZ regime of Fig. \ref{fig:ML}(a) we should expect a power law $M_{\|}g_{\|}^{-1/\delta}\propto (\hat\xi/\xi_{\|})^{\gamma/\nu}$ and, indeed, we can see that the exponent of this power law is close to $\gamma/\nu=1.75$. 

\begin{figure}[t]
\begin{center}
  \includegraphics[width=\columnwidth]{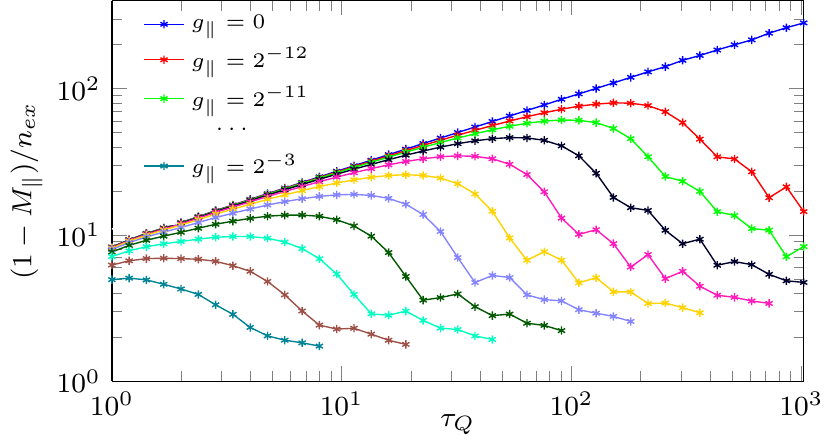}
\end{center}
  \caption{
  Here $M_{\|}$ and $n_{ex}$ are, respectively, the final magnetization and final density of kinks at $g_{\perp}=0$. $(1-M_{\|})$ is twice the density of spins pointing down, $n_{\downarrow}$.
  The plot shows the ratio $(1-M_{\|})/n_{ex}=2n_{\downarrow}/n_{ex}$ in function of $\tau_Q$ for 
  different biases $g_{\|}$. Here, we only show points where $n_{ex} > 10^{-5}$, similarly as in Fig.~\ref{fig:g0kinks}(a).
  } 
   \label{fig:cluster}
\end{figure}

The final $g_\perp=0$ is beyond applicability of the scaling hypothesis. Still, plots of $M_{\|}(g_\perp=0)$ in function of $\hat\xi/\xi_{\|}$ in Fig. \ref{fig:ML}(b) nearly collapse -- note slightly faster ordering for small fields than the one provided solely by the scaling hypothesis. The plot shows a crossover from the KZ to the adiabatic regime. In the latter there is almost full polarization, $M_{\|}(g_\perp=0)\approx1$. Almost all spins are pointing up except for a small fraction, $n_\downarrow=[1-M_{\|}(g_\perp=0)]/2$, pointing down. The down spins appear in clusters of adjacent sites. Each cluster is limited by an antikink and a kink on its left and right end, respectively. Figure \ref{fig:cluster} shows that in the adiabatic limit $n_\downarrow$ tends to twice the density of kinks $n_{ex}(g_\perp=0)$ implying the cluster size $1$, i.e., there are isolated down spins in the majority of spins pointing up. The cluster size increases with decreasing $\tau_Q$ until it reaches a maximum at the crossover to the KZ regime. 
In the KZ regime, where $n_{\downarrow}\approx1/2$ and $n_{ex}(g_\perp=0)\propto \hat\xi^{-1}$, the ratio $n_{\downarrow}/n_{ex}(g_\perp=0)\propto\sqrt{\tau_Q}$.  

{\it Conclusion.---}
We demonstrated that $\tau_Q g_{\|}^{(1+z\nu)/\beta\delta} \sim 1$ marks a crossover between the Kibble-Zurek and adiabatic regimes for, respectively, faster and slower quenches. Depending on where one is aiming, the crossover means that either a tiny symmetry-breaking bias is enough to make a ramp across a quantum critical point adiabatic or, equivalently, it shows how accurately the bias has to be tuned to zero in order to observe the KZ mechanism unperturbed. The final longitudinal magnetization crosses over from a full polarization deep in the adiabatic regime to a power law $M_{\|}\propto g_{\|} \tau_Q^{\gamma/(1+z\nu)}$ deep in the KZ regime.

\acknowledgments
{\it Note added. ---} After this work was completed, an experimental paper \cite{Volovik} appeared where analogous effect of an external bias is observed in a classical transition in superfluid $^3$He.

We thank Norman Yao for fruitful discussions at the initial stages of this project.
This research was funded by National Science Centre (NCN), Poland under Projects No. 2016/23/B/ST3/00830 (JD) and No. 2016/23/D/ST3/00384 (MMR), and by the U.S. Department of Energy at Los Alamos under the LDRD program. It was carried out with the equipment purchased thanks to the financial support of the European Regional Development Fund in the framework of the Polish Innovation Economy Operational Program (Contract No. POIG.02.01.00-12-023/08).

\section{Supplementary Material }
In this Supplementary Material we provide further details on the numerical simulations. 
The results presented in the main text were obtained by quenching the transverse field as
\begin{equation}
g_{\perp}(t) = 1-\frac{t}{\tau_Q}+\frac{4}{27}\left(\frac{t}{\tau_Q}\right)^3,
\label{rampS1}
\end{equation}
between initial $t_i = -\frac{3}{2} \tau_Q$ for which $g_{\perp}(t_i) = 2$, and final $t_f = \frac{3}{2} \tau_Q$ for which $g_{\perp}(t_f) = 0$. The qubic term in Eq.~\eqref{rampS1} is added
to start (and end) the quench smoothly with initial (and final) velocities $\dot{g}_{\perp}(t_i) = \dot{g}_{\perp}(t_f) = 0$. The critical point is reached for $t=0$. Close to the critical point,
to linear order we recover Eq.~\eqref{ramp} from the main text.

\begin{figure} [t]
\begin{center}
  \includegraphics[width=0.99\columnwidth]{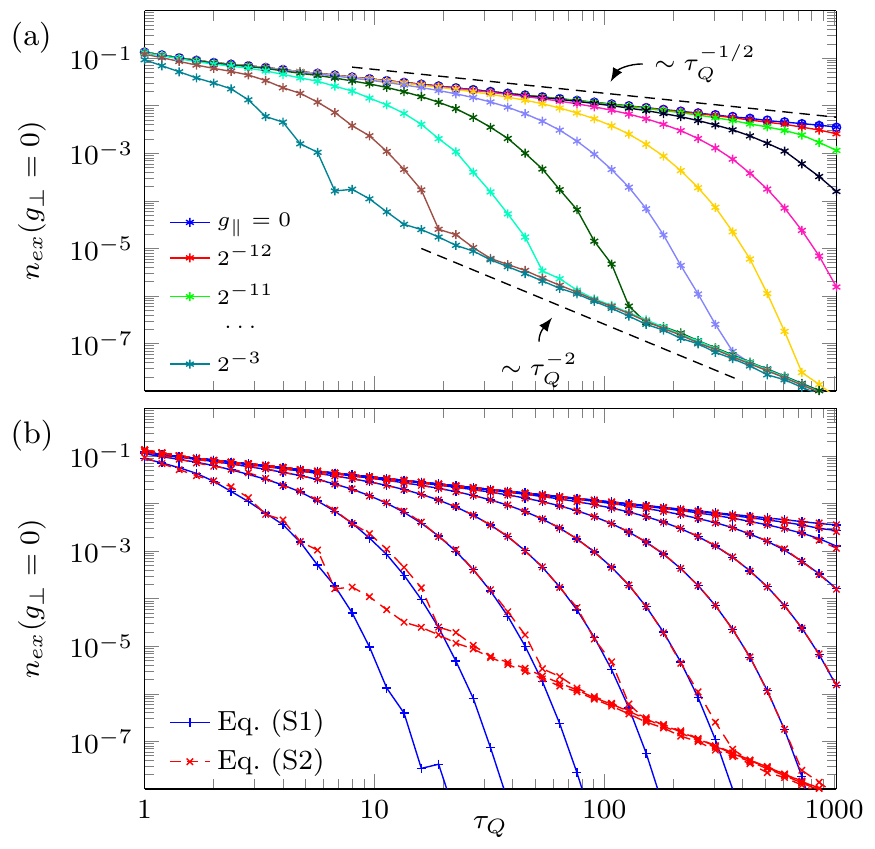}
  \end{center}
  \caption{
  In (a), the final density of kinks at the end of the linear ramp at $g_\perp = 0$, in the function of $\tau_Q$ for strictly linear protocol in Eq.~\eqref{rampS2}. Quench is started in the ground state at $g_i=4$.
  Non-vanishing velocity at the beginning of the quench leads to generation of excitations, which density
  vanishes as $\tau_Q^{-2}$. They become visible in the adiabatic limit of large bias and slow quench. 
  Circles show the exact analytically solution available for $g_\| = 0$ \cite{d2010}.
  In (b), we compare it with Fig.~\ref{fig:g0kinks}(a) of the main text where we use Eq.~\eqref{rampS1} with no discontinuity 
  of velocity at the beginning of the quench. }
 \label{fig:S1}
\end{figure}

We simulate the time evolution using uniform time dependent variational principle (uTDVP). The algorithm performing evolution of uniform matrix product state (uMPS) within this approach is masterfully explained in
Ref.~\cite{utdvp_review}. In the integration scheme we employ $4$-th order time-dependent Suzuki-Trotter decomposition \cite{Suzuki4}. Unlike the popular infinite time-evolving block decimation algorithm (iTEBD) \cite{vidal2007}, uTDVP benefits from inverse-free formulation in the thermodynamic limit. This has capital impact on numerical stability of the method and its ability to reliable simulate the quench evolution for large $\tau_Q$, as well as in the adiabatic limit. The results in the main text were obtained for uMPS bond dimension $D=64$ and the time-step $dt=1/8$.
We checked against smaller $D=32$ and $dt=1/16$ that those parameters are sufficient to obtain very good convergence.

In order to better demonstrate stability of the algorithm and the role of the cubic term in the quench protocol, we performed simulations with simple linear ramp 
\begin{equation}
g_{\perp}(t) = 1-\frac{t}{\tau_Q},
\label{rampS2}
\end{equation}
between initial $g_i = 4$ and final $g_f =0$. The resulting density of kinks is plotted in Fig.~\ref{fig:S1}a.
The linear ramp has a discontinuous time derivative $\dot g(t_i)$ because (implicitly) $g(t<t_i)=4$ is a constant. Due to this initial discontinuity, very deep in the adiabatic regime the kink density decays with $\tau_Q$ like $\tau_Q^{-2}$ rather than exponentially, see Fig.~\ref{fig:S1}a (for the same effect in the adiabatic limit for $g_\| = 0$ see e.g. discussion around Eq.~(48) in \cite{d2010}). In Fig.~\ref{fig:S1}b we directly compare those results with the ones for the smooth quench in Eq.~\eqref{rampS1} (from the main text). Apart from the above mentioned $\tau_Q^{-2}$ behavior in the adiabatic limit for purely linear quench, the results perfectly overlap. This shows that the qubic term in Eq.~\eqref{rampS1} does not change universal conclusions and scaling relations from the main text.

\end{document}